\documentclass[runningheads]{class1}
\usepackage{graphicx}
\usepackage{changepage}
\usepackage{xcolor}
\usepackage{soul,color}
\usepackage{color}
\usepackage{listings}
\usepackage{comment}
\usepackage{academicons}
\usepackage[utf8]{inputenc}
\usepackage{amssymb}
\usepackage{pifont}
\usepackage{longtable}
\usepackage{multirow}
\usepackage{xcolor,colortbl}
\definecolor{Gray}{gray}{0.85}
\usepackage[colorinlistoftodos]{todonotes}
\usepackage{academicons}
\usepackage{rotating}
\usepackage{tikzsymbols}
\usepackage{ragged2e}
\usepackage{array, threeparttable} 
\usepackage[T1]{fontenc}
\usepackage{booktabs}
\usepackage{caption} 
\begin{document}

\title{Provenance, Anonymisation and Data Environments: a Unifying Construction}
\titlerunning{Provenance, Anonymisation and Data Environments: a Unifying..}

\author {Muhammad Aslam Jarwar\inst{1}
\and
Adriane Chapman\inst{2}
\and 
Mark Elliot\inst{1} 
\and 
Fatemeh Raji\inst{2}
}
\authorrunning{Jarwar et al.}
%
\institute{University of Manchester, Manchester M13 9PL, UK \\
\email{\{aslam.jarwar, mark.elliot\}@manchester.ac.uk}\\
\and
University of Southampton, Southampton, SO17 1BJ, UK\\
\email{\{adriane.chapman, f.raji\}@soton.ac.uk}}
\maketitle              

\begin{abstract}
The Anonymisation Decision-making Framework (ADF) operationalizes the risk management of data exchange between organizations, referred to as "data environments". The second edition of ADF has increased its emphasis on modeling data flows, highlighting a potential new use of provenance information to support anonymisation decision-making. In this paper, we provide a use case that showcases this functionality more. Based on this use case, we identify how provenance information could be utilized within the ADF framework, and identify a currently un-met requirement which is the modeling of \textit{data environments}. We show how data environments can be implemented within the W3C PROV in four different ways. We analyze the costs and benefits of each approach, and consider another use case as a partial check for completeness. We then summarize our findings and suggest ways forward.

\end{abstract}

\setlength{\parindent}{0pt}
\setlength{\parskip}{1em}
\setcounter{footnote}{0}
\section{Introduction}
In the knowledge economy, large amounts of data are collected to support decision-making, policy analytics, service delivery etc. However, the usability of these data is constrained by the disclosure risks involved in data processing in general and data sharing in particular. One of the important tools used to mitigate this risk  is anonymisation.  The Anonymisation Decision-Marking Framework (ADF) operationalises the processes of functional anonymisation \cite{elliot2018functional}. 
This conceptualisation originated in the work of the \textit{data environment analysis service} \cite{elliot2010data}; a support system for the 2011 UK census focused on data confidentiality and  disclosure control \cite{willenborg2012elements,duncan2011concepts,hundepool2012SDC} and, in particular, re-identification risk assessment \cite {chen1998, skinner1998estimating,skinner2002measure}. The critical point underlying this concept is that disclosure risk resides not in the data themselves but in the relationship between the data and their environment.  Mackey and Elliot define the data environment as "the set of formal and informal structures, processes, mechanisms and agents that either: (i) act on data; (ii) provide interpretable context for those data or (iii) define, control and/ or interact with those data" \cite{mackey2013understanding}. 

Data environments come in a variety of types. For example, the open data environment, an end-user license management data environment, restricted access secure data environments etc. Notwithstanding this variety, the ADF framework assumes that all data environments can be described through four descriptive features: other data, agents, infrastructure, and governance. 

It follows from the foregoing that in order to apply the appropriate anonymisation processes, one needs to take account of both the data and their environment. Elliot et al. \cite{elliot2020anonymisation} developed the ADF to operationalise exactly such a process. The ADF emphasises that the appropriate anonymisation decisions for a given set of data are only be possible by considering the relationship between the data and their environment(s) which they call the \textit{data situation}.\\  

\vspace{-12pt}
\textbf{\textit{Problem statement}.} \textit{Data situations are often dynamic in that data move between environments for both processing and use. Thus, understanding contextual risk, and how to manage that risk through anonymisation, requires an awareness of, and capacity to map, the data flows between environments.}

Currently, capturing and mapping \textit{data situations} for analysis within the ADF framework is done manually, which is labor intensive and prone to errors. In order to automate this mapping, we propose the use of data provenance - a concept that is already mentioned in an informal sense in the ADF. By integrating provenance with the ADF, we will be able to track the flows of data and  recognise the upstream and downstream  data situations - both existing and proposed. We note that, data provenance has already been applied in the modelling of similar problems such as  situation awareness and decision making \cite{baclawski2017framework}, controlling of direct and indirect data flows \cite{rong2020provenance}, big data security and privacy \cite{gao2020big}. 

W3C PROV is a standard for provenance interoperability and for representing where data came from, and how it has been processed \cite{PROV-DM15,missier2013w3c}. PROV provides an abstract data model that includes agents, entities, activities, and relationship properties and which enables the representation of the  provenance of data and systems.

A critical element in the feasibility of linking provenance to the ADF is the representation of data environments. In the W3C PROV data model, two constructs \textit{bundles} and \textit{namespaces} might be considered to be candidates for such representation. In this paper, we examine the potential value of both of these solutions.  We also consider how the elements of PROV (i.e. Entity, Bundle, Agent, Activity) could be used to represent data environment features (agents, other data, infrastructure,  governance). We observe that there are limitations to representing data environments in this way and suggest some modifications which would enable full capture of the desired features.

The contributions of this work are as follows:
\vspace{-12pt}
\begin{enumerate}
    \item We outline -- using an ADF use case -- the requirements for provenance in the representation of data environments  (in section \ref{sec:usecase}).
    \item Using these requirements, we propose four different approaches to apply and extend W3C PROV to enable the representation of data environments for machine enabled reasoning (in section \ref{sec:impl}).
    \item We then analyse the four approaches (in section \ref{sec:analysis}) 
\end{enumerate}

\section{An ADF Use Case }\label{sec:usecase}
A seemingly simple data flow between environments can in fact be complex depending on the nature of the data and the environment(s), the intended data use and the responsibilities of the data situation's stakeholders. When data moves between environments (called a \textit{dynamic data situation} in ADF parlance), each environment produces a different risk profile, depending upon how the data interacts with the four defining features (other data, governance, infrastructure and agents). Below we describe an example use case. This example, drawn from \cite{elliot2020anonymisation}, is an idealisation of a common data situation; the sharing of data held by a national statistics agency with a research data service. 

\rule[3pt]{349pt}{1pt}

\parbox{12.5cm}{
\textbf{The set up:} the Government Office for National Data (GOND) collects several types of national level datasets. For example, national census data, public healthcare data, birth-death related data, pupil data from schools, traffic data from the smart cities sensors, etc.}
\vspace{-12pt}
\begin{itemize}
\item Part of GOND's remit is to make available some of those datasets for secondary research use. In service of this, it shares versions of the national datasets that it holds with the National Research Data Service (NRDS).  
\item The NRDS is part of University of Barsetshire. The NRDS's role is to acquire data from data holders, including GOND, under contract and then enable (and manage) access to those data under controlled conditions by researchers from research laboratories across the country.
\item GOND also releases highly aggregated data into the public domain (by definition an open environment).
\item The researchers carry out data analysis on GOND's data and then publish papers reporting on this analysis in the public domain. 
\item This data flow involves various loci of responsibility and control (key concepts in the ADF) over the data sharing in and from the different environments:
\begin{itemize}
\item GOND has \textit{indirect responsibility} and \textit{strategic control} over the data released from the NRDS environment into the open environment (in the form of analytical output within publications). GOND also has direct responsibility and control over the data released from its own environment into the public domain (in the form of aggregate statistics). 
\item NRDS's responsibility and control are different from GOND's, NRDS has \textit{direct responsibility} and \textit{operational control} over the data release from the output of publications.\footnote{See \cite{elliot2020anonymisation} for a more detailed discussion of the concepts of responsibility and control.}
\end{itemize}
\end{itemize}

\begin{figure}[hbt!]
\includegraphics[width=\linewidth]{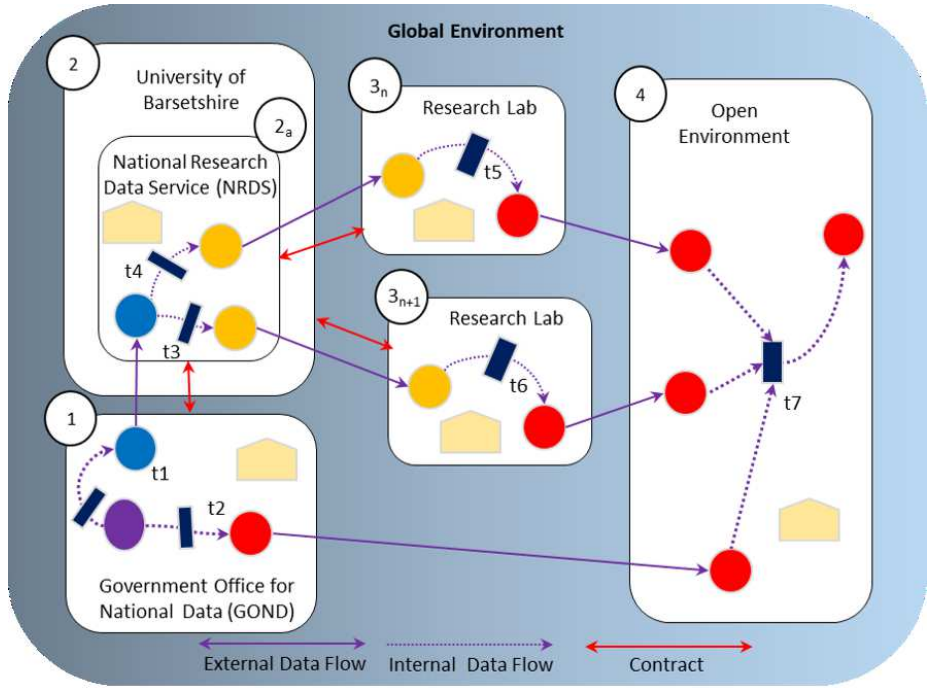}
\caption{A use case of data flows between and within multiple data environments. The red arrows indicate contractual agreements. The blue lines indicate data flow. Data environments are indicated by rounded rectangles, a circle represents a piece of data, a rectangle represent a process and a pentagon represents a user (in the data environment). The time for processing and sharing of data in the environments are labelled from \begin {math}t_1 \end{math} to \begin {math}t_7\end{math}.} \label{fig1}
\end{figure}

\rule[3pt]{349pt}{1pt}
The sketch diagram of this use case is shown in Figure \ref{fig1}. Four focal data environments are part of global data environment. GOND, the University of Barsetshire, and NRDS are represented as data environments 1, 2, and \begin{math}2_{a}\end{math}  respectively. The research labs and the open environment are labelled with data environments \begin{math} 3_{n} \end{math} to \begin{math}3_{n+1} \end{math} and 4 respectively. 
As shown in Figure \ref{fig1}:
\begin{enumerate}
    \item For the purposes of understanding this data situation the origin of the data flow is the GOND (1) data environment.\footnote{The same questions of granularity and scope affect the anonymisation use case as other uses of provenance information. In some instances, one may want to push the flow all the way back to the data subjects. For simplicity's sake here we are assuming that GOND are the origin.}. At $t_1$, the data are processed to make them compliant for sharing with (2), according to contractual obligations. At $t_2$, in parallel, the data are processed  more heavily for public release into the open environment (4).
    \item The data that is shared from GOND to the NRDS (2a), might be subjected to additional processing (disclosure controls) so that they can be shared with the various research labs who want to access the data for substantive analyses ($3_n,3_{n+1},....$).
    \item Each research lab analyses the data according to their particular needs and research questions. The research labs wish to produce publications and research datasets for public consumption (4).
    \item One of the goals of the ADF is to ensure that when data that has been derived from the same original data, are released by different organisations (or indeed at different times by the same organisation), inadvertent disclosures of personal information do not happen as a consequence. This is an increasingly critical issue which this data situation epitomises.
\end{enumerate}

\subsection{The Provenance Requirements of the ADF (using the GOND-NRDS use case)}
The next step in understanding the relationship between provenance information and anonymisation is to produce a set of representational requirements. Based on these requirements, a data environment formalism will be created using the W3C PROV data model (PROV-DM).\footnote{PROV-DM is the conceptual data model and core part of W3C PROV that defines each term used to represent provenance information \cite{moreau2015rationale}.} A specification of those requirements is as follows: 
\vspace{6mm}\\
\textit{\textbf{R1: The data environment construct}} \\ The data environment construct defines a boundary state that contains data. For example, GOND and NRDS are two closed data environments containing different data and within which different processing events occur.
\vspace{6mm}\\
\textit{\textbf{R2: Data environments within data environments}}\\ Sometimes an environment will contain other environments. For example, data flows between an organisation's sub-units for processing, auditing, etc. Another example is that the NRDS data environment is contained within the Barsetshire University data environment. In general, access control will be tighter in sub-environments than the host environment. \vspace{6mm}\\
\textit{\textbf{R3: Attaching attributes to data environments}}\\
To determine appropriate disclosure (control) practices, the purpose of data collection, type of data environment and any constraints and features (infrastructure and governance) of a data environment need to be recorded as attributes of that data environment. For instance,  GOND collects data from its partners for use and onward sharing via a legal gateway; the processing occurs in a restricted access data environment the parameters of which may be defined by - for example - a data sharing agreement, GONDS own data policies, the enabling legislation itself etc.  
\vspace{6mm}\\ 
\textit{\textbf{R4: Relationships between data environments}}\\
This describes the possible relationship from data environment to another data environment. For example, Within the NRDS, a research lab might have a specialised, secure processing environment which is owned and maintained by NRDS, but hosted for and used by the research lab. This is an example of a data environment with more complex relationships between data environment constructs than containment.
\vspace{6mm}\\  
\textit{\textbf{R5: Annotation of relational constructs}}\\
In order to reason over data environment interactions, controllers, processors, subjects, users, etc., it is important to allow the attachment of semantic meaning to the relationships between the constructs. For example, NRDS receive data from GOND and store it for onward sharing with researchers. In PROV, this might be achieved by labelling with \textit{prov:use} or \textit{prov:generated} but these labels do not represent all of the required information needed for the ADF. 
\vspace{6mm}\\
\textit{\textbf{R6: Representation of agents, data and processes within a data environment}}\\
Agents might include data controllers, data processors, data users and data subjects. Data includes datasets, reports, etc. Processes include data extractions, sharing, storing, sampling aggregating, etc. In our use case the research labs contain agents (users), a process (analysis), input data for the analysis and output data (e.g. tables, models, graphs).   
\vspace{6mm}\\
\textbf{\textit{R7: Data governance instruments: contracts}}\\
There are numerous types of data governance instruments that affect what can and can't be done with data. One important type is  
the contract; typically a data sharing agreement to share, exchange and use data between the environments.   For example, in our use case  GOND share data with NRDS based on the  contract between them.  
\vspace{6mm}\\ 
\textbf{\textit{R8: Access and control (direct and indirect)}}\\
A record of the access and control mechanisms over the data and services. For instance, GOND has a data dissemination function that can be used by NRDS (based on some contract). GOND also has indirect control over data releases from the NRDS environment (in that the output disclosure control policy of NRDS will be defined by GOND).   
\\  
\section{Supporting Data Environments with W3C PROV}
\label{sec:impl}
In this section, we will explain how PROV can be applied to support the data environment representation requirements outlined in section \ref{sec:usecase}.  We will show that the existing W3C PROV data model  does already support some of the data environment representation requirements in that some of them can be mapped onto PROV elements. However, there are some data environment specific requirements that need extensions in PROV. We will describe four possible mechanisms: namespaces with or without supporting structures and bundles with or without an extension.

\subsection{Namespaces and support structures} \label{subsec:namespaces}
The namespace concept was inspired by the World Wide Web architecture and was designed to make objects interoperable across technologies and platforms \cite{moreau2015rationale}. In PROV-DM, Namespaces are a Uniform Resource Identifier (URI);
a provenance graph can contain multiple - 
possibly many - namespaces. The namespace is a candidate for use as an identifier to capture the idea of multiple data environments (including data environments within data environments) and their associated entities, activities, agents, etc. By using Namespaces and prefixes we could differentiate the representation of nested data environments and can access related elements information through namespace concatenating and de-concatenating. 

\begin{figure}[hbt!]
\includegraphics[width=\textwidth]{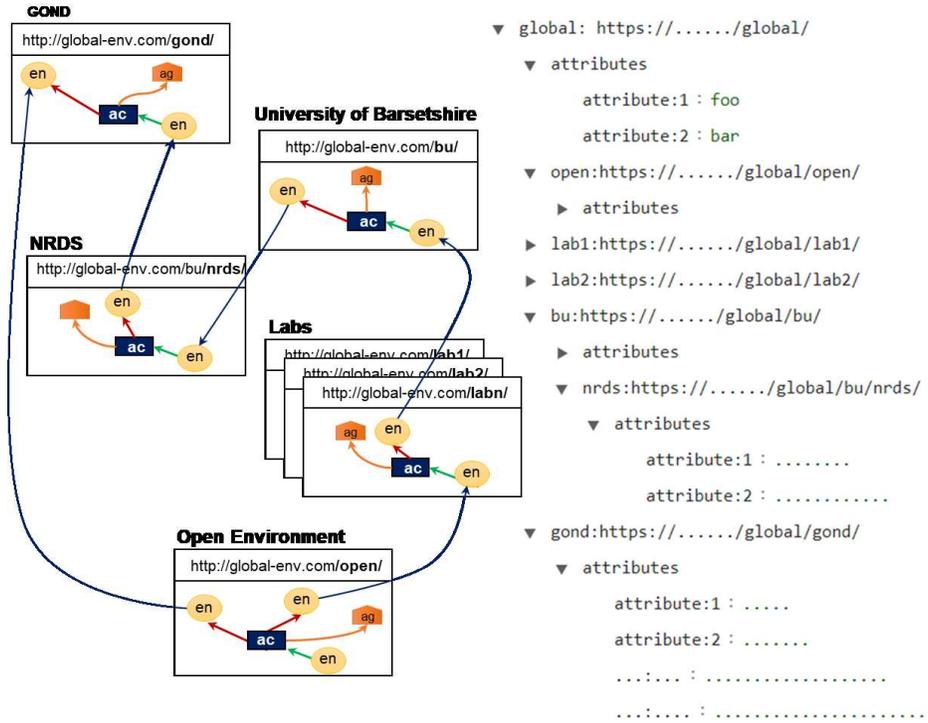}
\caption{Illustration of the use of namespaces to represent Data Environments: ag, ac, and en indicates agent, activity, and entity respectively; the right-hand part  shows data environments with attribute attachment using namespaces. Relationships across namespaces could be captured in the same manner.} \label{fig:namespaces}
\end{figure}

For example, we might refer the University  of Barsetshire and NRDS data environments as \textit{http://global-env.com/bu/ 
} and  \textit{http://global-env.com/bu/nrds/ 
} respectively. (Note: In the example use case, the NRDS data environment is a part of University of Barsetshire environment). We can also express the control mechanism over the data environments and its elements with namespace features. The visual representation of the GOND-NRDS use case  with the support of Namespaces and PROV constructs is shown in Figure \ref{fig:namespaces}. 
 
In Figure \ref{fig:namespaces}, there are  five main data environments with separate namespace. For instance, the GOND data environment can be recognised with namespace \textit{http://global-env.com/gond/}. The elements of GOND such as entity\_001 can be accessed with \textit{http://global-env.com/gond/entity\_001\#}. Similarly the agent with an id "agent\_controller\_001" from NRDS data environment can be recognised with a \textit{http://global-env.com/bu/bu/nrds/agent\_controller\_001\#}. Additionally, as illustrated  Figure \ref{fig:namespaces}, the data provenance for research labs can be tracked in forward and backward called as forward and backward chaining. The forward chaining informs how the research labs data will be utilised and backward chaining tracks the sources of data and the contracts between research labs with the data providers. Moreover, the right hand side of Figure \ref{fig:namespaces}  shows the pseudo code of attributes attachment with the data environment through namespaces' support. In the pseudo code, the global data environment has two attributes with values \textit{foo} and \textit{bar}.

While namespaces have potential for representing the bounded nature of data environments, and what has occurred within a given data environment and it's sub-environments, namespaces alone are not enough to satisfy all of the requirements identified in section \ref{sec:usecase}. For instance, the attachment of additional attributes to the data environment itself and contracts between the data environments  cannot be accommodated. Additionally, relationships among namespaces beyond containment cannot be captured. For instance, it is possible in namespaces to distinguish that \textit{ http://www.nytimes.com}  data environment that contains a sub-data environments related to advertising functions, \textit{http://www.nytimes.com/ads}. However, within our use case, there is more than strict-hierarchical containment. For example, researchers from one of the Research Labs might have a specialised data analysis environment built-by, hosted-by and managed-by NRDS, but considered an enclave of both NRDS and the Research Lab. In this case, namespaces do not capture enough information to represent this relationship.

To solve these issues, an additional set of structures would need to be created. For instance, a separate document which extends namespaces and allows attachment of attributes, could be used.

\subsection{Bundles and Extended Bundles }
\begin{figure}[hbt!]
\includegraphics[width=\linewidth, height=8cm]{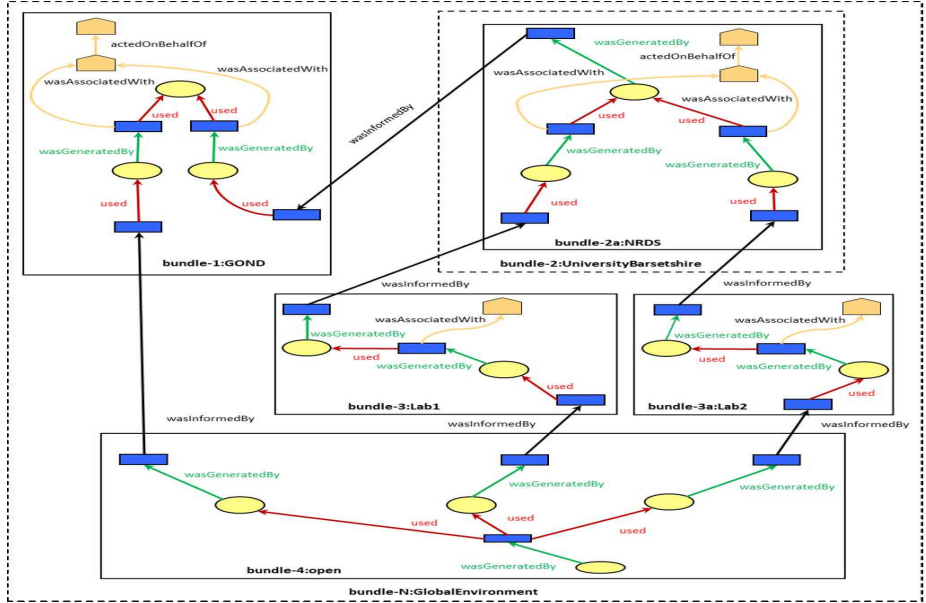}
\caption{A representation of the GOND-NRDS use case supported with PROV bundles. Please note that the nested data environments are shown with dotted lines are for illustration of use case and currently these are not supported in PROV.}  \label{fig:bundle}
\end{figure}

In PROV, the bundle concept has some similarities to the data environment construct. 
The bundle  is itself an entity which provides provenance information regarding the creation and modification of a group of entities \cite{mckenna2019modelling}. 
For example, a bundle can contain the entities, activities, agents, and the relationships between them.  
Within a given bundle, the  data, and data processes can be represented with entities and activities respectively. Bundles can also support entities with attributes. This can help us to add necessary metadata to the entities. The excerpt view of the GOND-NRDS use case representation as supported by PROV bundles is shown in Figure \ref {fig:bundle}.

In Figure \ref {fig:bundle}, the large rectangles delineate data environments each represented as a PROV bundle. Each bundle contains  data environment elements (represented as nodes) and  relationships between those elements (represented as edges). For example,  in the "bundle-1:GOND" data environment, the processes (small blue  rectangles)  are using a piece of data for generating another piece of data. For these processes a data processor (agent expressed with pentagon) is responsible (the responsibility relationship is shown with "wasAssociatedWith"). The relationship between the data controller and data processor\footnote{The terms data processor and controller are key terms in the General Data Protection Regulation (GDPR). See https://ico.org.uk/for-organisations/guide-to-data-protection/guide-to-the-general-data-protection-regulation-gdpr/controllers-and-processors/what-are-controllers-and-processors/ for definitions.} is shown with  "actedOnBehalf" property.

The data flow between one data environment and another environment (we can say that from bundle to bundle) is shown  with "wasInfromed" property. For instance the data flows in direction of bundle-1:GOND->bundle-2a:NRDS-> (bundle-3:Lab1, bundle-3:Lab2), etc is represented with "wasInfromed" property. 

We can also see in Figure \ref {fig:bundle} that the NRDS data environment ("bundle-2a:NRDS") is a sub environment of University of Barsetshire (bundle-2:UniversityBarsetshire). We note that in ADF terms, NRDS is said to have \textit{direct control} over the labs environment for releasing of data, whereas GOND has \textit{indirect control}. To support the representation of control (and its companion concept of responsibility) would need additional mechanisms to be added to PROV but this lies outside of the immediate scope of this paper.     
         
W3C PROV constructs were designed to be extensible \cite{moreau2015rationale}. In previous work, PROV has been extended to express the provenance of big data security supervision \cite{gao2020big}, provenance access control \cite{missier2020abstracting}, data privacy protection based on GDPR using provenance \cite{davari2019access} and  managing mutable entities by adding reference derivations and checkpoints \cite{pimentel2018versioned}.
Likewise, we can extend the existing structure of PROV bundles in order to support and express the requirements of ADF with more flexibility. For example, by extending we can attach  additional metadata to the bundle construct, which would enable us to define the different types of data environments. Another extension that we would need in PROV Bundles, is support for nested data environments. 


\section{Comparative Analysis} \label{sec:analysis}

\subsection{Requirements Completeness}

\begin{figure}[hbt!]
    \includegraphics[width=\textwidth]{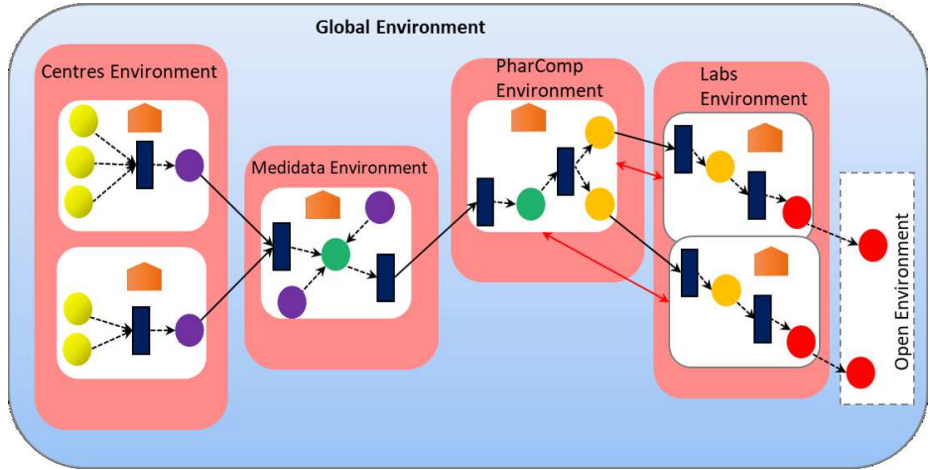}
    \caption{A Second ADF use case with data environments: The red arrows indicate contractual agreements. The black lines indicate data flow. Data environments are indicated by rounded rectangles, A circle represents a piece of data, a rectangle represent a process and a pentagon represents a user in the respective data environment}
    \label{fig:usecase2}
\end{figure}

To test the specification of the provenance needed for the ADF define in section \ref{sec:usecase}, we analyse a second ADF use-case (please refer Figure \ref{fig:usecase2})  in which:
\begin{itemize}
\item Data from clinical trials are generated at several participating centres.
\item The data are uploaded electronically by the participating centres to a company called Capturedata which offers an electronic data capture and management system for the pharmaceutical industry. 
\item PharComp, a European pharmaceutical business, extracts and downloads the clinical trial data from the Capturedata database onto PharComp systems for analysis. 
\item PharComp shares some of the data with researchers, for use in public health research. 
\item Researchers publish their analysis in journal articles in the public domain. These data will not include information that directly identifies the patients, and additional steps are taken to safeguard the patients’ confidentiality.
\item  Explicit consent has been given by trial participants for secondary research using of anonymised versions of their data.
\end{itemize}

We use this use case to confirm that the requirements identified in Section \ref{sec:usecase} are correct (they are sufficient in description to cover equivalent elements of this use case) and complete (there are no additional requirements identified in this use case).

The \textbf{data environment construct} is required to fully capture the data situation as encompassing the environments of collection centers, Capturedata, PharComp, research labs and the open environment. The data collection centers and the research labs contain sub-environments for various types of data collection, processing and to meet research protocol requirements, confirming the \textbf{“data environments within data environments”} requirement. In this use case, the Capturedata and PharComp are types of restricted access data environments and can be accessible to only authorised users and researchers. To express this type of and restriction to these data environments we need to \textbf{represent data, agents and processes}.  

As with the GOND-NRDS use case, here we need to \textbf{annotate the relationship constructs} between the data environments of collection centers and Capturedata, where the data is collected and stored instead of data derivation/usage or generation. This will support the semantic meaning of relationship construct in the data provenance. Given the nature of the data collected for the pharmaceutical company, contracts specifying data collection, exchange and control exist (\textbf{contracts}). PharComp has  indirect control over the data release in the open environment in the form of publications and the researcher must follow the code and conduct given by the PharComp, this requirement can be represented with \textbf{access and control} category as described in section \ref{sec:usecase}. 

One of the requirements presented in section \ref{sec:usecase} the \textbf{relationship between data environments} is not found in this use case, as there are no data environments with multiple institutional ownership and use. However, if the specific example contained an enclave in PharComp in which regulatory employees from a government could review specific data, this requirement would be needed. There are no additional requirements that seem necessary to capture the provenance of data situation for the purposes of the ADF.

\subsection{Analysis of Implementation Approaches}
Table \ref{tab2} shows the data environment representation requirements outlined in section \ref{sec:usecase} and the ability of each of the implementation options discussed in Section \ref{sec:impl} to meet those requirements. 

\begin{table}[!htbp]
 \caption{Use case requirements analysis (NB: here Namespace + includes attributes and PROV constructs}
  \label{tab2}
  \begin{tabular}{|p{5cm}|c|c|c|c|}
    \hline
    \multirow{2}{*}{Representation requirements} &
    \multicolumn{4}{c|}{Support} \\  
      & Bundle & Namespace & Namespace+  &  Bundles+\\
    \hline
    Data Environment Construct  & \checkmark & \checkmark & \checkmark & \checkmark \\
     \hline
    Data Environments within Data Environments  &  - & \checkmark & \checkmark &\checkmark \\
     \hline
    Attaching Attributes to Data Environments & - & - &  \checkmark &\checkmark \\
     \hline
     Relationship between Data Environments & \checkmark & - & \checkmark &\checkmark \\
     \hline
     Annotation to   relational constructs  & - & - & \checkmark &\checkmark \\
     

     \hline
     Representation of agents, data and processes within DE  & \checkmark &  \checkmark & \checkmark &\checkmark \\
     \hline
     Data governance instruments: contracts  & - &  - & \checkmark &\checkmark \\
    \hline
     
     Access and control & \checkmark & \checkmark & \checkmark &\checkmark \\
     
    \hline
  \end{tabular}
 
\end{table}

The nesting of data environment (i.e. data environments within data  environments) is one of the important features.  However, with bundles we cannot represent  nested data environments because PROV does not allow the nesting of bundles  \cite{moreau2015rationale}.This gap is one of the drivers for bundles+. This requirement is  supported by namespaces (and so namespaces+ can also support nesting). 

The ability to attach attributes to a data environment is also an important feature for the ADF. Neither bundles or namespaces support attachment of attributes. For example, currently, we cannot express following using PROV bundles:
\begin{adjustwidth}{4em}{4em}
\textit{bundle (EX-A:GOND, \\ prov:envType ="Government",\\ prov:governance-accessType="Restricted",\\
 prov:governance-userdefinition="TrainedLevel2",\\
 prov:infrastructure="ISO27001")}.\\
\end{adjustwidth}
The additional structures provided in Namespace+ do allow attributes to be maintained with the namespace information. However, Bundles+ is a more elegant option; using the W3C PROVs standard of attaching attributes to other object types, but expanding that notion to bundles+. 

As we observed in the GOND-NRDS use case (see Figure \ref{fig1}), the GOND data environment contains the representation of collected, processed and shared data along with the data processes, agents, and contracts (i.e. contract with the NRDS), and IT infrastructure and services. In order to create the provenance graph for this data situation, the relationships between these elements would need to be supported with PROV properties. For example  \textit{ wasGeneratedBy (entity\_id , activity\_id)}, and \textit{used(activity\_id, entity\_id)} properties could be used to represent the relationship between the GOND collected data and processing of the data to generate the new dataset for NRDS. 

Both the bundles and namespaces solutions could naturally support the representation of agents, processes and entities using native W3C PROV concepts. On the other hand, supporting additional metadata such as annotation with the relationship constructs is not fully supported in PROV. However, this  could be managed  by attaching additional attributes with the relationship construct (this approach was used  by  \cite{pimentel2018versioned} for the purpose of tracking changes in entities over time). Attaching annotation will also be helpful here in selecting an appropriate disclosure control processes. Bundles+ supports this requirement. The GOND-NRDS contracts are  supported by both Bundles+  and Namespaces+. The representation of access control requirement is supported by all four constructs.  

\subsection{Validation of Data Environment Representation}
Currently, the source code for PROV validation is not openly available. However, the source code for the SEIS-PROV’s\footnote{SEIS-PROV is a domain specific extension based on the W3C PROV data model, used in the seismological data processing. This extension defines a new namespace with entities, activities and attributes in the context of seismology.}   document validation is available at \cite{SEISPROV}. The SEIS-PROV validation mechanism is implemented in python. Using this validation tool as an exemplar, to validate the representation that includes the data environments within data environments feature, the PROV document should include a formalism for data environments:
\begin{math}
[d_i=I_i \cup d_1,....,[d_n=I_n \cup  d_{n-1}] 
\end{math}
where  n is number of data environments and PROV elements instances, I is the top level prov element instance and d is the data environment instance.The value of i will be between 0 and n. 

The PROV validation mechanism has two components: inference and constraints. The inference  component deals with the fixing of missing information based on the definition of the element defined in the PROV data model. The constraint component includes a checking mechanism that deals with uniqueness, ordering, impossibility and typing. Impossibility checks for prohibited patterns, while the typing constraints check the type of identifier when it is used in relations.    
Inferencing should be performed over the document, and the elements should be categorised as per the definition. 
For example, similar entities in two different data environments might be categorised according to the prefixes of definition or prefixes over the data environment.
     
\subsection{Translation and Visualisation} 
In order to share the data environment representation with other stakeholders we may need to support the translation from PROV-N to other formats (e.g. json, provx, turtle, trig, svg, rdf, xml) and vice versa. Therefore, to accommodate the proposed extension to PROV bundles, the existing translation and visualisation mechanism would also need to be updated.

Our goal would be to incorporate the support for the data environment representation the in PROV python implementation. The PROV python implementation provides a PROV serialisation module \cite{provpythonpkg} that provides various classes to transform PROV document from one format to another format. For example, \textit{ProvJSONSerializer, ProvRDFSerializer, ProvNSerializer, ProvXMLSerializer} provides the implementation to translate PROV in JSON, RDF, prov notation and XML formats respectively. All of these serialisation classes would need consequential changes to support the bundle extension. 

For graphical visualisation of provenance statements, the PROV python implementation uses three open source libraries pydot \cite{pydot2021}, Graphviz \cite{ellson2001graphviz}, and DOT language \cite{gansner2006drawing} in \textit{prov.dot} module. The \textit{prov.dot} module also needs substantial changes in \textit{prov\_to\_dot()},  \textit{ \_bundle\_to\_dot()},      \textit{\_attach\_attribute \_annotation()},  etc
methods that translate the  provenance statements into visualisation graphs. 
These methods would also need updates along with additional methods to support the graphical visualisation of  the complexity of data situations.

\section{Related Work} \label{sec:relwork}
W3C PROV has been used elsewhere to capture provenance for the protection of data subjects' confidentiality, and the security of data. 

A W3C PROV based provenance model has been proposed by Benjamin et al. \cite{ujcich2018provenance} that uses the PROV data model ontology and data protection ontology to express the provenance for compliance with the European Union (EU) General Data Protection Regulation (GDPR). The Agent, Activity, and Entity classes from the PROV ontology were extended with sub-classes to express the provenance of GDPR compliance. For example, \textit{Subject}, \textit{Controller}, \textit{Processor}, and \textit{Supervising-Authority} sub-classes were introduced within the agent class. The Activity class was extended with  two additional sub-classes: \textit{Process} and \textit{Justify}. Similarly, the Entity class was extended with three sub-classes:  \textit{PersonalData}, \textit{Request} and \textit{Justification}. The relationships among the classes were expressed with PROV properties. Both of the ADF examples presented in this work fall under GDPR regulations, and the extensions introduced in Benjamin et al. \cite{ujcich2018provenance} would facilitate some of the more general requirements of \textbf{representation of agents, data and processes} and \textbf{contracts} within data environments.

To support  provenance of mutable values by time-versioning entities, a PROV extension has been developed by adding the reference sharing and checkpoints feature \cite{pimentel2018versioned}. These features were built on top of PROV events that track a version of and object or entity through change or generation events (i.e. \textit{prov:Generation}) and access or usage events (i.e. \textit{prov:Usage}). The checkpoint attributes were used with the PROV entities, activities, relationship properties for tagging and tracking of changes in the entities over the time period. For this purpose, two namespaces (i.e. version and script) were created to support the checkpoints mechanism. These were used for both general PROV extension concepts and specific script concepts. However, this approach increases the overhead for querying the provenance graph due to folding and unfolding for adding the checkpoints.  

The PROV data model has also been extended with new relationship properties in order to supervise the security of data streaming \cite{gao2020big}. These extensions focus on collecting the provenance information about data operations inside and outside of big data clusters; representing the data interaction flow between the clusters. The harvested relationship provenance information of the graph is analyzed for the detection of anomalies in the data. The anomaly detection and reasoning mechanism checks for inconsistency between the nodes and edges. 

Pahl et al. have used the PROV data model along with blockchain technology to implement a trust analysis platform. PROV was used to capture the features for verifying the originality and source of data received from sensors and edge cloud devices.\cite{pahl2018architecture}, 

Finally, PROV has been used to protect data provenance content that is sensitive and subject to disclosure control \cite{missier2020abstracting}, modelling the threat of attack to supply chain electronic management system \cite{halak2021cist}, and detection of bottlenecks in the system by analyzing the patterns in the provenance graph \cite{boutamina2018bottleneck}.

\section{Conclusions and Future Work} \label{sec:concl}

In this paper, we have considered a new application of provenance: to support anonymisation of data exchanged across organisations and environments. To this end, we introduce the Anonymisation Decision Framework (ADF) which is used to reason about data flows and anonymisation. Through analysis of the ADF, how it is applied, and the information required to make such decisions, we have identified how provenance might be utilised more formally.

In order to do this effectively, we need to be able to represent one of the core components of the ADF approach the \textit{data environment}, an organising concept constituted from other data, agents, governance processes and infrastructure. We identified the key properties of such environments from an idealisation of a  real world use case which can be mapped with W3C PROV elements: entities, bundles, activities, and agents. 

We analysed how data environments can be represented within the W3C PROV. We observed that, in order to fully express the features of data environments, the existing PROV constructs are not sufficient and would need extending. We identified four different candidate mechanisms within the W3C PROV, and evaluated each with respect to trade-offs of cost, maintenance and suitability for the problem. While two obviously do not pass muster, the other two are viable solutions, with one \textit{Namespaces+} that utilises existing W3C PROV structures but requires an additional management, and the  second \textit{Bundles+} which requires an extension to PROV.
\bibliographystyle{plain}

\end{document}